\documentclass[twocolumn,showpacs,amsmath,amssymb]{revtex4}
\usepackage{epsfig}
\usepackage{amsmath}
\usepackage{amssymb}
\usepackage{graphics}
\usepackage{xcolor,soul}
\usepackage[colorlinks=true,citecolor=blue,linkcolor=blue]{hyperref}

\DeclareMathAlphabet{\bi}{OML}{cmm}{b}{it}

\begin{document}
\title{Hierarchical nanostructuring approaches for thermoelectric materials\\ 
with high power factors}
\author{Vassilios Vargiamidis}
\email{V.Vargiamidis@warwick.ac.uk}

\author{Neophytos Neophytou}

\affiliation{School of Engineering, University of Warwick, Coventry, CV4 7AL, UK}%
\begin{abstract}
The thermoelectric power factor of hierarchically nanostructured materials is investigated using the non-equilibrium Green's function method for quantum transport, including interactions of electrons with acoustic and optical phonons. We describe hierarchical nanostructuring by superlattice-like potential barriers/wells, combined with quantum dot barriers/wells nanoinclusions as well as voids in the intermediate region. We show that these structures can be designed in a way that the power factor is not only largely immune to the presence of the nanostructure features, but under certain conditions benefits can be achieved as well. Interestingly, we show that these design approaches are linked to the energy relaxation of the current flow and whether charge carrier scattering is limited by elastic or inelastic processes. In particular, when nanostructures form potential barriers, the power factor can be substantially enhanced under elastic scattering conditions, irrespective of nanostructuring density and potential barrier heights. When inelastic scattering processes dominate, however, the power factor is inevitably degraded. In the case in which nanostructures form potential wells, despite a slight decrease, the power factor is quite resilient under either elastic or inelastic scattering processes. These nanostructuring design approaches could help open the path to the optimization of new generation nanostructured thermoelectric materials by not only targeting reductions in thermal conductivity, but simultaneous improvements in the power factor as well.
\end{abstract}
 \maketitle

\section{Introduction}

Thermoelectric (TE) materials convert heat from temperature gradients into electricity and vice versa. Their performance is quantified by the dimensionless figure of merit $ZT = \sigma S^2 T / (\kappa_{\text{e}} + \kappa_{\ell})$ where $\sigma$ is the electrical conductivity, $S$ is the Seebeck coefficient, $T$ is the operating temperature, $\kappa_e$ is the electronic thermal conductivity, and $\kappa_\ell$ is the lattice thermal conductivity. The product $\sigma S^2$ is known as the power factor ($PF$). Traditional TE materials, which are mostly semiconductor doped alloys of Sb and Bi$_2$Te$_3$ at room temperature, and PbTe or SiGe at higher temperatures, reach $ZT \approx 1$. Over the last several years, however, numerous other materials have been explored, such as transition-metal dichalcogenides (TMDC) \cite{GDing16,HHuang16,WHuang14}, skutterudites \cite{YTang15ncom,YTang15nmat}, phonon-glass-electron crystal structures \cite{Beek15nmat}, half-Heuslers \cite{Stern16ntech,CFu15ncom}, oxides \cite{YYin17}, etc. A large number of these materials demonstrate $ZT$ above $1$, primarily due to the reduction of their thermal conductivity \cite{Beretta19}.

In order to achieve even further reductions in thermal conductivity, the majority of these materials are explored in the context of nanostructuring. Many approaches towards this route are common practice, such as superlattice-like geometries \cite{Mizuno15}, alloying \cite{XWWang08}, heavy doping \cite{Ikeda10}, nanoporous materials \cite{Verdier16,RYang05,JHLee08}, nanograining \cite{Neo13ntech,Bennett16}, nanoinclusions (NIs) \cite{Biswas12,Gayner16,Zou15,Hopkins11,Vineis10,Popescu09}, etc. Nanoinclusions, in particular, cause scattering of short wavelength phonons with mean-free-paths in the order of nanometers, which otherwise contribute significantly in the thermal conductivity of common TE materials, such as in PbTe \cite{Bo11}. This technique is applied to a broad range of materials, including Bi$_2$Te$_3$ \cite{Fan11,Keshavarz14}, PbTe \cite{Heremans05,Biswas12,Hsu04}, SiGe \cite{Ahmad16,XWWang08,Zhu09}, MnSi \cite{Saleemi15}, and SnTe \cite{Tan14}, to name a few. 

Furthermore, undoubtedly, one of the most successful approaches to reduce thermal conductivity is hierarchical nanostructuring, where distortion features are placed within a matrix material at the mesoscale (grains/boundaries), microscale (NIs) and nanoscale (atomic defects). These scatter phonons of various wavelengths and reduce phonon transport across the entire spectrum. Indeed, by nanostructuring PbTe in a hierarchical manner, record high value of $ZT=2.2$ was achieved due to drastic reductions in $\kappa_\ell$ \cite{Biswas12}. More recent works have achieved even higher $ZT$ up to $2.5$ at $923$ K \cite{GTan16}. For such a success, the thermal conductivities in these materials reached values well below the amorphous limit which is $1-2$ W/mK at room temperature \cite{Kearney18}, and thus, cannot be further reduced easily. Therefore, further benefits to $ZT$ can only be achieved from the enhancement of the $PF$. In the majority of cases, however, nanostucturing degrades the electrical conductivity and the $PF$ as well. In addition, the adverse interdependence of the electrical conductivity and Seebeck coefficient does not allow flexibility in $PF$ improvements.

The importance of retaining high $PF$s has recently become more appreciated in nanostructured TE materials. Reference \cite{Zhao14}, in particular, emphasizes the importance of matrix/inclusion band alignment to retain the original conductivity of the material and to avoid degradation in the $PF$. Large $PF$ improvements were also demonstrated in highly doped nanocrystalline Si \cite{Neo13ntech}, especially in the presence of nanovoids \cite{Lorenzi14}. While the impact of nanostructuring on the thermal conductivity can be more clearly understood as a general increase in phonon scattering, the same cannot be assumed for the $PF$. Due to the complexity in geometry, theoretical works to date (by us and others) focus on one type of nanostructured feature at a time, i.e. only superlattices (SLs) \cite{Thes16JEM}, only nanocrystalline boundaries \cite{Kearney18}, only NIs \cite{Fan11,Ahmad16,Liu12,Foster17,Peng14,Zhou11}, etc. In several cases, the conclusions vary substantially, from reports of large $PF$ benefits to only moderate or none; a consequence of the difficulty in accurately simulating and optimizing the complexity of geometries coupled with the complexities of the nanoscale transport physics in the presence of various types of nanostructured features. Indeed, the complexity of the electronic transport, combining semiclassical effects, quantum effects (i.e. quantization, tunneling, interferences, resonances), ballistic and diffusive regimes, as well as the geometry details with multiple features and feature types, makes accurate modelling a difficult task. Thus, it is imperative to shed more light and establish a high level of understanding of the $PF$ behavior in the presence of more than one nanoscale feature type, both qualitatively and quantitatively, if $ZT$ is to be maximized. 

In this work we use the non-equilibrium Green’s Function (NEGF) method to calculate the electron transport properties of two-dimensional (2D) nanostructures when superlattice-type boundaries and NIs are present simultaneously (see Fig.~1). NEGF provides a unified, geometry flexible, fully quantum mechanical simulation approach well-suited for this problem. We present a systematic investigation of how such complex geometries affect the $PF$ in the cases where the nanofeatures impose potential barriers, or potential wells for charge carriers. We explore the influence of the heights of those barriers, as well as their number density. We then identify the design approaches that will allow for $PF$ immunity in hierarchically nanostructured materials, and in some cases, even significant improvements.

The paper is organized as follows. In Sec.~II we describe our NEGF approach including our calibration procedure and indicate the geometries we study. In Sec. III we present and discuss our results, in Sec. IV we present a discussion on optimal nanostructuring, and finally, in Sec. V we conclude.

\begin{figure}[t]
\vspace{-0.1in}
\hspace*{-0cm}
\includegraphics[width=7cm,height=15.2cm]{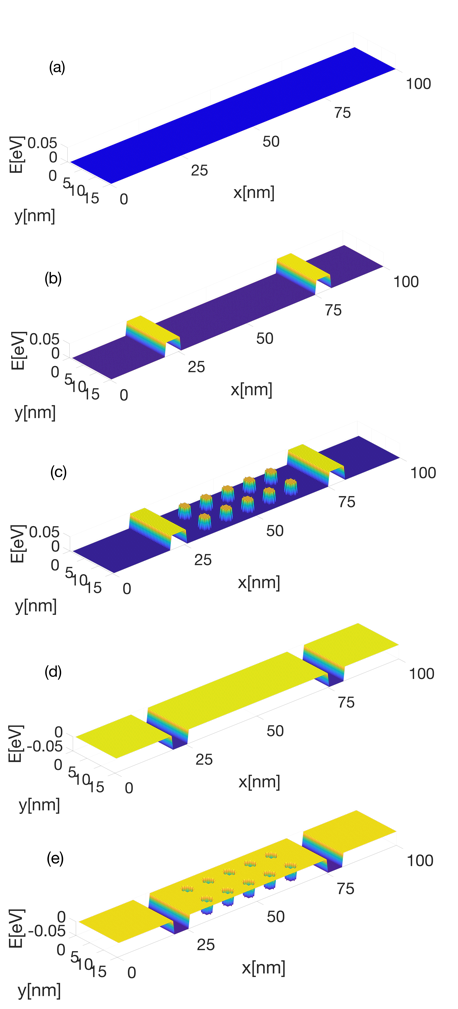}\\
\vspace*{0.2cm} \caption{\label{fig1} (Colour online)
Geometries of the hierarchical nanostructured materials that we consider in this paper. (a) The pristine channel. (b) Channel with SL-type barriers. (c) Channel with SL barriers and quantum-dot potential barrier NIs. (d) Channel with SL-type wells. (e) Channel with SL-type wells and quantum-dot potential well NIs.
}
\end{figure}

\section{Approach}

We employ a 2D quantum transport simulator based on the NEGF formalism including electron-phonon (e-ph) scattering in the self-consistent Born approximation. We include both scattering of electrons with acoustic phonons (elastic scattering) and with optical phonons (inelastic scattering). The formalism and the details of the specific 2D simulator, which we developed in order to capture phonon scattering, as well as its convergence details, are described in several works of ours and of others \cite{Koswatta07,Datta05,Anantram08, Foster17,Var17pssa}. However, in order to be able to better discuss certain characteristics of e-ph scattering in the context of this work, we include here a brief description of the model with notation adopted to our system.

In the NEGF method a system/channel, described by a Hamiltonian $H$, is connected to two contacts (left and right), which are represented by self-energy functions $\Sigma_{\text{L}}$ and $\Sigma_{\text{R}}$. The Hamiltonian is constructed using a 2D effective mass (single-orbital) tight-binding grid uniformly spaced in the $x$- and $y$-directions, resulting in a banded matrix. The 2D channels we consider within the effective mass approximation, have a uniform $m^\ast = m_0$ in the entire channel, where $m_0$ is the rest mass of the electron. Thus, we do not consider a specific material, rather our study aims in providing first-order qualitative guidance into the design of high power factor nanostructured materials. These self-energies represent the influence of the semi-infinite Left and Right leads on the channel, respectively. Note that $\Sigma_{\text{L}}$ and $\Sigma_{\text{R}}$ are energy dependent, and non-Hermitian. They are formed using the first/last channel layers from which electrons are injected into the channel (and thus, they have the same size as these layers), and are calculated using the Sancho-Rubio iterative scheme \cite{Sancho_Rubio_85}. They are added to the first/last layer elements of the channel Hamiltonian. The e-ph scattering process in the device enters the NEGF formalism through the self-energy function $\Sigma_{\text{S}}$. One can view the scattering process as just another contact described by $\Sigma_{\text{S}}$, similar to the actual contacts described by $\Sigma_{\text{L}}$ and $\Sigma_{\text{R}}$, however, $\Sigma_{\text{S}}$  is added to all diagonal elements of the Hamiltonian.

The retarded Green's function for the device is given by \cite{Datta05}
\begin{equation}
G(E) = \left[ \left( E + i \eta^+ \right) I - H - \Sigma(E) \right]^{-1} ,
\label{eq1}%
\end{equation}
where $\eta^+$ is an infinitesimal positive number which pushes the poles of $G$ to the lower half plane in complex energy, $I$ is the identity matrix, and $\Sigma(E)$ is the sum of the self-energies
\begin{equation}
\Sigma(E) = \Sigma_{\text{L}}(E) + \Sigma_{\text{R}}(E) + \Sigma_{\text{S}}(E) .
\label{eq2}%
\end{equation}
It proves useful and convenient to define the in-scattering self-energies due to contacts as:
\begin{equation}
\Sigma_{\text{L, R}}^{\text{in}} (E) = - 2 \text{Im}  \left[ \Sigma_{\text{L, R}} (E) \right] f_{\text{L, R}} (E) ,
\label{eq3}%
\end{equation}
where $\text{Im}[...]$ is the imaginary part and $f_{\text{L, R}}$ is the Fermi distribution for the left and right leads. Similarly, the out-scattering self-energies are defined as:
\begin{equation}
\Sigma_{\text{L, R}}^{\text{out}} (E) = - 2 \text{Im}  \left[ \Sigma_{\text{L, R}} (E) \right]  \left[ 1 - f_{\text{L, R}} (E) \right] .
\label{eq4}%
\end{equation}
With $\Sigma_{\text{L, R}}^{\text{in/out}} (E)$ one can express the electron and hole correlation functions as:
\begin{equation}
G^{\text{n}} (E) = G(E) \Sigma_{\text{L, R}}^{\text{in}} (E) G^\dagger (E) ,
\label{eq5}%
\end{equation}
\begin{equation}
G^{\text{p}} (E) = G(E) \Sigma_{\text{L, R}}^{\text{out}} (E) G^\dagger (E) .
\label{eq6}%
\end{equation}

Assuming that the system consists of 2D grid/lattice points with uniform spacing $a$, and making the nearest neighbour tight-binding approximation, the current density between grid points $j$ and $j+1$ is given by
\begin{eqnarray}
\nonumber J_{j, j+1} = \frac{i e}{\hbar}
\\* &&\hspace*{-1in} \times (2)  \int_{-\infty}^{\infty} \frac{d E}{2 \pi} \left[ H_{j+1,j} G_{j, j+1}^{\text{n}} (E) - H_{j,j+1} G_{j+1, j}^{\text{n}} (E) \right]   ,
\label{eq7}%
\end{eqnarray}
where $H_{j+1, j} = H_{j, j+1}^\dagger$ are the hopping matrix elements of the Hamiltonian, and $(2)$ is for the two spin directions.

A second source for in-scattering and out-scattering of electrons from an occupied state is the e-ph interaction. The self-energy at point $j$ and energy $E$ has two terms corresponding to scattering from $(j, E+\hbar \omega)$ and $(j, E-\hbar \omega)$. Within the Born approximation the in-scattering self-energy into a fully empty state is \cite{Mahan87}
\begin{equation}
\Sigma_{\text{S}}^{\text{in}} (E) = D_{0} [ n_{\text{B}} G^{\text{n}} (E-\hbar \omega) + (n_{\text{B}} + 1) G^{\text{n}} ( E + \hbar \omega) ]  .
\label{eq8}%
\end{equation}
where $D_0$ represents the e-ph scattering strength at grid point $j$, $n_{\text{B}}$ is the Bose-Einstein distribution function for phonons of energy $\hbar \omega$, and $G^{n} (E-\hbar \omega)$ is the electron density at $E-\hbar \omega$. The first and second terms in Eq.~(\ref{eq8}) represent in-scattering of electrons from $E - \hbar \omega$ (phonon absorption) and $E + \hbar \omega$ (phonon emission) to $E$, respectively. The out-scattering self energy, $\Sigma_{\text{S}}^{\text{out}} (E)$, from a fully filled state at energy $E$ is given by \cite{Mahan87}
\begin{equation}
\Sigma_{\text{S}}^{\text{out}} (E) = D_{0} [ (n_{\text{B}} + 1) G^{\text{p}} (E-\hbar \omega) + n_{\text{B}} G^{\text{p}} ( E + \hbar \omega) ]  ,
\label{eq9}%
\end{equation} 
where $G^{\text{p}} (E-\hbar \omega) $ and $G^{\text{p}} (E+\hbar \omega)$ are the densities of unoccupied states at $E-\hbar \omega$ and $E+\hbar \omega$. The first and second terms in Eq.~(\ref{eq9}) represent out-scattering of electrons from $E$ to $E-\hbar \omega$ (phonon emission) and $E+\hbar \omega$ (phonon absorption), respectively. In the case of acoustic phonons, $\hbar \omega \rightarrow 0$, and so in Eqs.~(\ref{eq8}) and (\ref{eq9}) we let $D_0 n_{\text{B}} \rightarrow D_{\text{AP}}$ (making use of the commonly employed equipartition approximation – see details in Appendix A) while in the case of optical phonons $D_0 \rightarrow D_{\text{OP}}$, which are taken to be constant throughout the channel.

The strength of the phonon scattering is adjusted such that the mean-free-path of electrons is $\lambda = 15$ nm. The way we calibrate this, is that we initially simulate a channel with length $L = 15$ nm in the ballistic regime, and then we increase the electron - acoustic phonon scattering strength $D_{\text{AP}}$ in the NEGF formalism, until the channel conductance drops to 50\% of its ballistic value ($D_{\text{AP}}=0.0026$ eV$^2$). This effectively fixes a mean-free-path of $15$ nm for the channel (under acoustic phonon scattering conditions alone), a value that is comparable to that of common semiconductors such as silicon \cite{Neo11PRB}. The nanostructured geometries that we consider are shown in Fig.~1. The channels have length $L=100$ nm and width $W=15$ nm. Thus, with a mean-free-path of $\lambda = 15$ nm, the channel we consider is long enough for the transport to be diffusive. Note that the computational cost of NEGF simulations scales with the third power in channel width, and thus we only consider narrow channels of widths as given above. However, using wider structures should not affect our final conclusions. Throughout the paper we assume room temperature $T = 300$ K.

In the case that we consider electron-optical phonon scattering only, we use an optical phonon energy of $\hbar\omega = 0.06$ eV (a value similar to that of silicon longitudinal optical phonons, see for example \cite{Jacoboni83}), and simply use the same value of the scattering strength in the simulations, i.e., $D_0 = D_{\text{OP}}=D_{\text{AP}}=0.0026$ eV$^2$. This gives an energy relaxation length of $\lambda_{\text{E}} = 13$ nm, which, in the well region of length $60$ nm, dictates semi-relaxation of the electron energy, after they pass over the SL barriers. We consider this semi-relaxation of energy because it has been shown that it provides optimal conditions for the $PF$ of SL materials \cite{Neo13ntech,Thes15JAP,Kim09}. Note, however, that in this way the conductance of the channel with only optical phonon scattering will be in general larger compared to that with only acoustic phonon scattering. This is due to the fact that electron scattering with optical phonons has weaker phonon absorption rates, as a consequence of lower than unity phonon occupation number. Thus, we cannot compare the two cases quantitatively. In the case where we consider both acoustic and optical phonon scattering, we simply divide the scattering strengths $D_{\text{AP}}$ and $D_{\text{OP}}$ by a factor of two, which (as we show later) gives similar conductances for their pristine channels as in the acoustic phonon scattering case.

In order to calculate the power factor $G S^2$, where $G$ is the channel conductance and $S$ the Seebeck coefficient, we use the fact that the Seebeck coefficient is the average energy of the current flow \cite{Kim11} (see Appendix B for derivation details)
\begin{equation}
S = \frac{1}{q T L} \int_{0}^{L} \langle E(x) - E_F \rangle dx ,
\label{eq13}%
\end{equation}
where $q$ is the carrier charge ($q=-\vert e \vert$ for electrons and $q=\vert e \vert$ for holes) and $\langle E ( x ) \rangle$ is the energy of the current flow along the transport direction, defined as:
\begin{equation}
\langle E ( x ) \rangle = \frac{\int I_{ch} (E, x) E dE}{\int I_{ch} ( E, x) dE}  ,
\label{eq14}%
\end{equation}
where $I_{ch}(E,x)$ is the energy and position resolved current. This is the quantity we integrate in order to get in Eq.~(\ref{eq7}) the total current as $J = \int I_{ch}(E) dE$. Note that the current is constant along the channel at each cross section, however, its energy is not constant, i.e. the charge carriers can gain or lose energy as they propagate. This happens in the presence of inelastic scattering (optical phonons). Thus, the energy of the current is position dependent and it is its scaled integral in the channel which provides the overall value of the Seebeck coefficient. In experimental settings, one extracts the Seebeck coefficient from the open circuit voltage upon the application of a thermal gradient along the channel, as $S = \Delta V / \Delta T$, which equivalently can also be computed by $S = I_{ch (\Delta V = 0)} / G \Delta T$. In \cite{Kim11}, it was validated that the two methods of extracting the Seebeck coefficient are equivalent, which makes it easier in time consuming simulations (as the ones we undertake) to only run the $\Delta V \neq 0$ case and still be able to extract the Seebeck coefficient by integrating the energy of the current flow over the length of the channel, and that is how we extract the Seebeck coefficient in this work.

Note that from NEGF we obtain the conductance $G$, rather than the conductivity $\sigma$, because NEGF simulates a 2D channel with specific dimensions, thus, the units of the two quantities are also different. This could be converted to conductivity using the channel dimensions, however since we do not have a specific thickness associated with our 2D simulation ($W=15$ nm, $L=100$ nm only), we use the conductance $G$ from here on. Also note that in all our results below we refer to $G$ as the conductance, not to be confused with the Green’s function in Eqs.~(\ref{eq1})-(\ref{eq9}), for which it is also customary to use $G$.

Figure 1(a) shows the pristine channel that we begin with. The conduction band reference is set at $E_{\text{C}} = 0$ eV. The Fermi level is placed also at $E_{\text{F}} = 0$ eV as this provides the highest $PF$ \cite{Kim09,Thes15JAP,Foster17}. Thus, we begin with an optimal channel as our basis, and we then proceed with inserting nanostructured features. The first nanostructured feature is SL type boundaries that form potential barriers for electrons as shown in Fig. 1(b). We consider thickness of $L_{\text{SL}} = 5$ nm and arbitrarily choose barrier heights of $V_{\text{SL}} = 0.05$ eV $(\approx 2 k_B T)$. We then add NIs with diameter $d=3$ nm each in between those boundaries as indicated in Fig. 1(c). In these structures, the Fermi level is placed at $E_{\text{F}} = V_{\text{SL}} = 0.05$ eV, so that the carriers are allowed easily to flow over the SL barriers. The NIs are modeled as potential barriers of cylindrical shape in rectangular arrangements within the matrix material, and their number density and barrier heights are varied, as discussed in the text below. The choice of the SL and NI sizes are such as to minimize the influence of quantum tunneling, which becomes strong and detrimental for the $PF$ for feature sizes below $3$ nm \cite{Thes15JAP}. We also consider nanostructured geometries where the potential barriers are replaced with potential wells (Figs. 1(d)-(e)). Finally, we also consider the situation in which the NIs are replaced by voids (not shown). The NEGF approach is ideal for such geometries as they can be described precisely when the Hamiltonian of the system is constructed.
\begin{figure}[!htbp]
\vspace{-0.1in}
\hspace*{0.2cm}
\includegraphics[width=8.6cm,height=14.2cm]{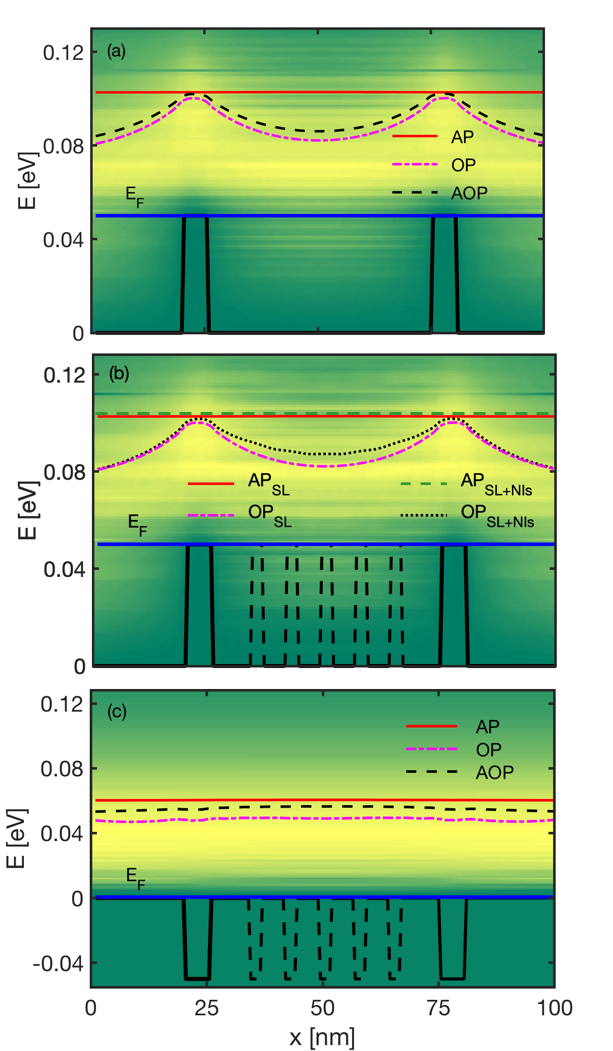}\\
\vspace*{0.2cm} \caption{\label{fig2} (Colour online)
Average energy of the current flow $\langle E ( x ) \rangle$ as defined in Eq.~(\ref{eq14}) along the channel length with (a) SL barriers, (b) SL barriers and NI barriers, and (c) SL wells and NI wells. The black lines (solid and dashed) represent the potential barriers. The yellow-green colormap shows the energy resolved current $I_{ch}(E,x)$, with yellow corresponding to regions of large flow in the case of OP scattering. The blue lines represent the position of the Fermi level $E_F$. In (a) the red-solid line shows $\langle E ( x ) \rangle$ in the presence of AP only (i.e., elastic scattering), in which case charge carriers have constant energy above the barriers/wells. The magenta-dashed-dotted line shows $\langle E ( x ) \rangle$ in the presence of OP only (i.e., inelastic scattering), in which case carriers relax their energy in the regions between the barriers and absorb phonons to gain energy and overpass the barriers. The black-dashed line shows $\langle E ( x ) \rangle$ in the presence of AOP. In (b) the red-solid and magenta-dashed-dotted lines are the same as in (a) for the SL structure. The green-dashed and black-dotted lines show $\langle E ( x ) \rangle$ in the presence of AP and OP, respectively, for the SL+NIs structure. In (c) the red-solid line shows $\langle E ( x ) \rangle$ in the presence of AP, the magenta-dashed-dotted for OP, and the black-dashed for AOP.
}
\end{figure} 

It turns out that a lot about the thermoelectric transport can be understood by looking at the energy of the current flow, $\langle E(x) \rangle$, along the transport direction, as defined in Eq.~(\ref{eq14}). This states, as expected, that the higher in energy the current flows with respect to the Fermi level, the higher the Seebeck coefficient is. It also provides some indication about the electrical conductivity, i.e. the higher the energy of the current flow, the more electrons with higher velocities are utilized (assuming no complex bandstructure effects cause velocity reductions), and the higher the conductivity could be. As we will show below, it proves to be a very useful feature in understanding TE transport particularly in the nanostructured materials we consider, and below we describe how we use it to interpret our simulation results.

Figure 2 shows $\langle E ( x ) \rangle$ for different structure cases and scattering conditions. Figure 2(a) considers a channel with two SL-type barriers inserted (black SL barrier lines). The colormap in all subfigures shows the energy and spatial regions where the current flows (yellow) in the case where both acoustic and optical phonon (AOP) scattering are taken into account, i.e., elastic and inelastic scattering, respectively. In this situation, $\langle E ( x ) \rangle$ is indicated by the dashed-black line. Clearly, electrons absorb optical phonons, overpass the potential barriers and then they relax into the wells by emitting optical phonons. The solid-red line shows $\langle E ( x ) \rangle$ under purely acoustic phonon (AP) scattering-limited conditions. In this case scattering is elastic, and the energy of the current is constant along the channel. The dashed-dotted-magenta line shows the optical phonon (OP) scattering-limited transport case, where the energy of the current flow is now slightly lower and the degree of energy relaxation slightly larger compared to the AOP case. Note that we do not compare the two cases on equal basis as the scattering rates for the OP case and AOP case are different, i.e. the OP scattering strength is halved in the AOP case, so relaxation is weaker. 

Figure 2(b) shows $\langle E ( x ) \rangle$ in the case where NIs are placed in between the SL barriers (dashed line barriers). The horizontal dashed-green line shows again that under acoustic (elastic) scattering conditions $\langle E ( x ) \rangle$ remains constant along the channel length, and is almost identical to the SL case (solid-red line). In the case of optical (inelastic) scattering conditions (dotted-black line), $\langle E ( x ) \rangle$ is relaxing in the region between the SLs, as expected. However, the presence of NIs reduces the number of available states that carriers can fall into after emitting phonons of energy $\hbar \omega$. In addition, since NIs disturb the low energy electrons, the average energy of the current $\langle E ( x ) \rangle$ slightly increases compared to the SL alone case \cite{Foster17}. Consequently, the degree of energy relaxation in the presence of NIs (dotted-black line) is smaller than that in the absence of NIs (dashed-dotted magenta line-repeated here from Fig.~2(a)), signaling a higher Seebeck coefficient. In Fig. 2(c), we show the corresponding case where the SL regions and the NIs form potential wells, and thus, less obstruction of transport is expected. Indeed, in this case $\langle E ( x ) \rangle$ is uniform throughout the channel, not only in the case of elastic scattering, but also in the case of inelastic scattering, as the wells are too narrow for the electrons to relax into.
\begin{figure}[t]
\vspace{-0in}
\hspace*{-0cm}
\includegraphics[width=8.4cm,height=14cm]{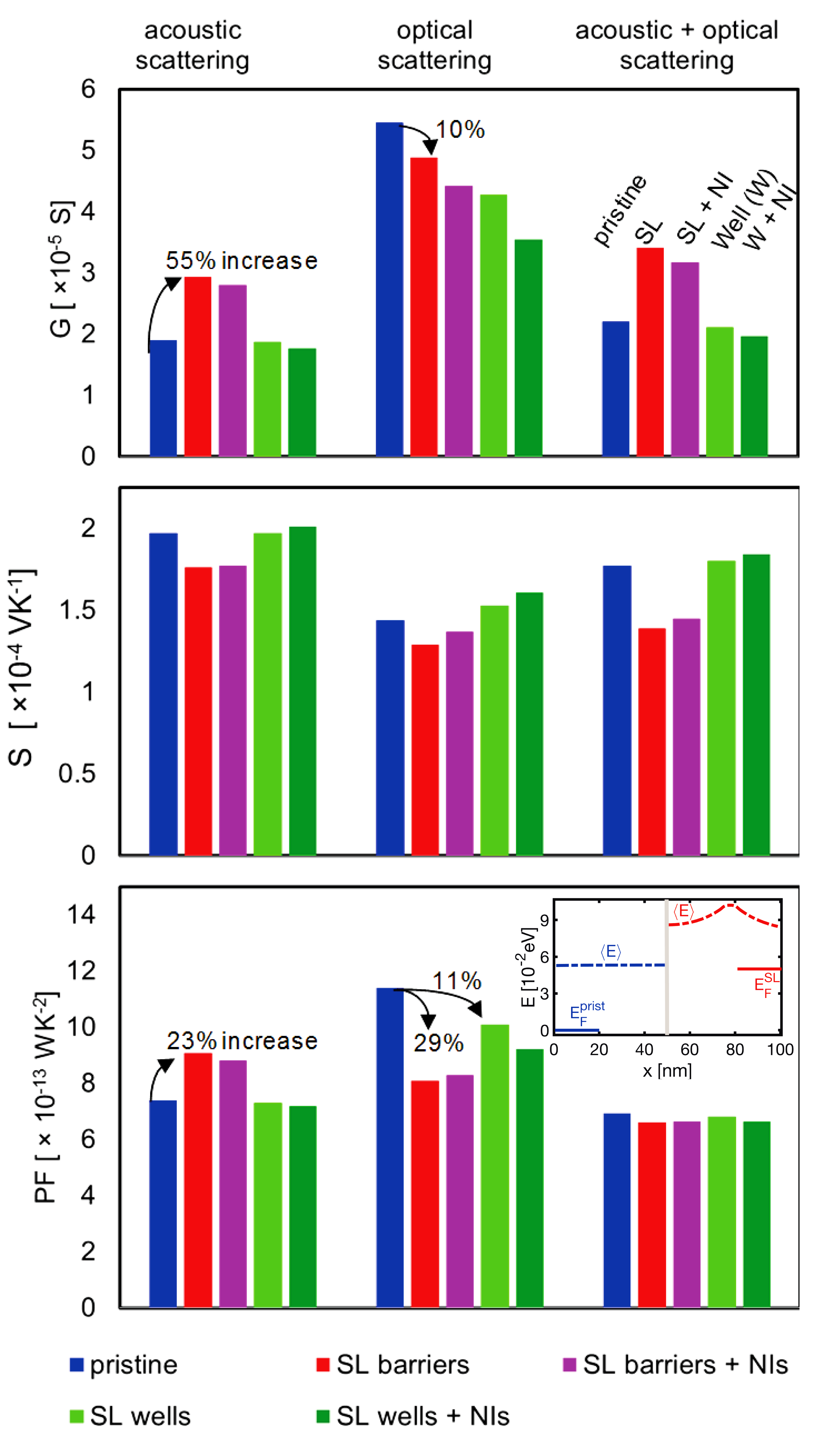}\\
\vspace*{0.2cm} \caption{\label{fig3} (Colour online)
Row-wise, summary of conductance $G$, Seebeck coefficient $S$, and power factor $PF$, for key structure examples as shown in Fig. 1. The blue bar indicates the pristine channel (Fig. 1a), the red bars the SL barrier channels (Fig. 1b), the magenta bars the SL+NI barrier channel (Fig. 1c), the light green bar the SL well channel (Fig. 1d), and the green bar the SL+NI well structure (Fig. 1e). Column-wise, the three different groups are results for AP scattering only, OP scattering only, and AOP scattering. Notice that the $PF$ in the SL case is higher than that of the pristine channel by $23\%$ under elastic AP scattering conditions only. This is retained when NIs are inserted. 
}
\end{figure}
\begin{figure*}
\hspace{-0.75cm}
  \begin{minipage}[c]{0.71\textwidth}
    \includegraphics[width=\textwidth]{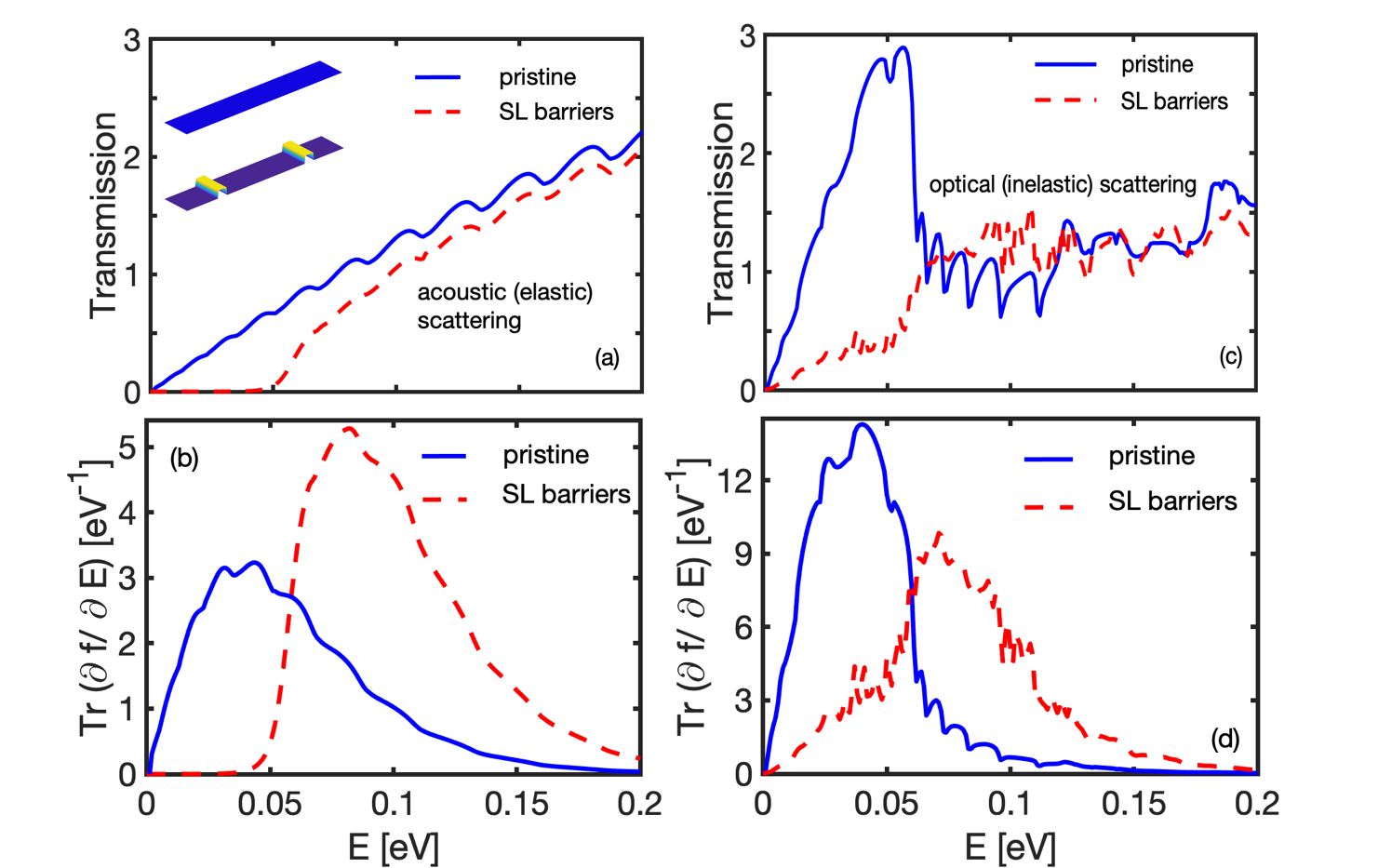}
  \end{minipage}\hfill
  \begin{minipage}[c]{0.3\textwidth}
    \caption{
       \label{fig4} (Colour online)
(a) and (b): Transmission $Tr$ and $Tr \left( \partial f / \partial E \right)$ versus electron energy $E$ for the pristine channel (solid-blue lines) and for the channel with SL barriers (dashed-red lines) in the AP scattering regime. (c) and (d): The same as in (a) and (b), but in the OP scattering regime. The Fermi level is $E_{\text{F}} = 0$ eV in the pristine channel and $E_{\text{F}} = 0.05$ eV in the channel with SL barriers. Notice that the transmission for the channel with SL barriers increases faster than that for the pristine channel for energies close to $E_{\text{F}}$ in the elastic scattering regime. This is a consequence of the fact that charge carriers at higher energies have higher velocities which, therefore, leads to higher conductivity.
    } \label{fig:03-03}
  \end{minipage}
\end{figure*}

\section{Results and discussion}

In the basis of the above observations, we will explain the $PF$ behavior in these nanostructures. It turns out that the dominance of elastic or inelastic scattering processes has a significant effect in the $PF$ \cite{Kim11,Thes16JEM}, since relaxation lowers $\langle E ( x ) \rangle$ and degrades the Seebeck coefficient. Therefore, for each one of the five geometries shown in Fig. 1, we consider AP scattering alone, OP scattering alone, and finally both AOP scattering combined. We begin by presenting in Figs. 3(a)-(c) some key simulation results for those basic structures with respect to the electronic conductivity, Seebeck coefficient, and $PF$, respectively. In the case of the SL barrier structures, we align the Fermi level with $V_{\text{SL}}$ at $0.05$ eV for optimal performance, whereas in the other cases with the band edge $E_{\text{C}}$. The different bars correspond to the geometries of Fig. 1 as follows: i) blue bars-pristine structure, ii) red bars-SL barrier structure, iii) magenta bars-SL barriers plus NIs structure, iv) light green bars-SL wells structure, and v) green bars-SL wells with NIs structure. The three different column groups show the corresponding values for AP scattering, OP scattering, and AOP scattering, respectively. 

\textbf{Elastic scattering improves the optimal PF:} We first focus on AP (elastic) scattering conditions, where the results for $G$, $S$, and $PF$ are shown in the first column group of Fig. 3. We compare how each quantity changes with respect to the pristine material (blue bars) for each nanostructured geometry. Interestingly, by raising $E_{\text{F}}$ high in the bands, the conductance $G$ (first column, first row, red bar) is increased by $55\%$, despite the introduction of the SL barriers (we discuss the reasons behind this behavior below). The Seebeck coefficient naturally drops (red bar in second row, first column), but overall the $PF$ is \textit{increased} in the SL structure by $23\%$ compared to the pristine structure. This indicates that the energy filtering, provided by potential barriers that cut lower parts of the Fermi distribution, is more effective at degenerate conditions as long as energy does not relax, as also pointed out in earlier studies \cite{Kim09,Neo13ntech,Thes15JAP}. However, the reason the $PF$ improves originates from the significant increase of the conductivity, rather than from the Seebeck coefficient which is actually degraded. The introduction of NIs in the region between the SL barriers (magenta bars), has a small degrading effect in the electrical conductance of the channel, as the NIs introduce additional scattering. However, this reduction is not strong, and the $PF$ is retained at values higher than those of the pristine material by $19\%$. Thus, we demonstrate here that it is indeed possible to achieve significant $PF$ improvements (rather than reductions) in a hierarchically nanostructured material. This $PF$ improvement combined with the expected very low thermal conductivity, can lead to high $ZT$.

It is important, thus, to clarify the reason behind the increase in $G$, which is responsible for this $PF$ improvement. The physical origin for this behavior lies in the fact that the charge carriers propagate on average at higher energies, which allow higher group velocities, and thus higher mobility. This is evident in Fig. 4(a), which shows the transmission versus energy of the pristine material (solid-blue line) and of the SL material (dashed-red line), extracted at every $x$-point as $Tr(x) = \left( h / e^2 \right) \left( I(x) /(f_1 - f_2) \right)$. Note that $I(x)$ is constant in the case of AP which makes the $Tr(x)$ constant, but not in the case of OP and AOP. In the semiclassical Boltzmann transport formalism, the transmission is related to the transport distribution function (TD) $\Xi ( E )$ via $Tr_n (E) = \left( W / L \right) 2\pi\hbar \Xi_n (E)$, where $\Xi_n(E) = g_n(E) \upsilon_n^2 (E) \tau_n (E)$ is the TD function per subband $n$ \cite{NeoNanoLett,Neo11PRB,Mahan96PNAS, Scheidemantel03}. In the usual case where $\tau_n (E) \propto 1/g_n(E)$, then $\Xi_n(E) \propto \upsilon_n^2 (E) \propto 2 E / m^\star$, which is linear in energy as we also observe within NEGF in Fig. 4(a). In the case of the SL, the transmission opens up for energies above the SL barrier at $V_{\text{SL}} = 0.05$ eV, but when this happens, the slope is \textit{larger} compared to that of the pristine material. This indicates that carriers, after passing over the SL barriers, relax on the higher velocity states of the intermediate region, and propagate with larger group velocities. Figure 4(b) shows the transmission scaled over the derivative of the Fermi function $Tr \left( \partial f / \partial E \right)$, which captures the part of the transmission that actually contributes to transport. Clearly, the higher peak in the case of the SL structure (dashed-red line), indicates larger conductivity. Thus, the deeper the well between the SL barriers, the higher the mobility and conductivity of the material.

\textbf{Inelastic scattering degrades the power factor:} What actually causes reduction in the $PF$, is the presence of energy relaxation, which is a result of inelastic scattering, in our case OP scattering. In the second column of Fig. 3, we show how the TE coefficients change in the presence of inelastic scattering alone. Because the phonon absorption process is weaker in the OP case as a consequence of the lower than unity phonon occupation number, the conductance of the pristine channel case (blue bars) is larger compared to that of the AP only case. Thus, we do not intent to quantitatively compare the two cases anyway. Furthermore, it is clearly seen that both the conductance $G$ and the Seebeck coefficient $S$ are reduced in the SL structure, $G$ even more in the SL plus NIs structure (compare the blue bar to the red and magenta bars in the second column group of Fig. 3). This reduction in both $G$ and $S$ leads to a large reduction in the $PF$ by $29\%$. As shown in Fig. 2 above, the energy relaxation process of charge carriers in the region between the SL barriers causes electrons to propagate at lower velocity states, which leads to reduction of the conductivity and of the Seebeck coefficient. This is again shown clearly in Figs. 4(c) and 4(d), where we plot the transmission $Tr$ and $Tr \left( \partial f / \partial E \right)$ for the pristine (blue lines) and the SL structure (red-dashed lines) when only OP scattering is considered. The peak in the pristine case at low energies is a consequence of not having phonon emission processes for carriers with energies smaller than the OP energies considered here $\hbar \omega = 0.06$ eV. In the case of the SL structure, however, where the $E_{\text{F}}$ is raised at the $V_{\text{SL}}$ level, emission is actually possible. In this case the scattering rate into lower energy states is larger, leading to reduction of the electronic conductivity compared to the pristine channel, despite the higher carrier energies and velocities. (Note that in the presence of inelastic scattering, the current flow, although constant along the channel, varies in energy. Thus, the transmission versus energy function is also spatially varying. In this case we still have transmission for energies below $V_{\text{SL}}$  because we extract the transmission at a point in the middle of the channel, where relaxation allows current flow at lower energies).  

In the case where we introduce NIs in the region between the SL barriers (Fig. 3, column 2, magenta lines), the conductance suffers even more. However, the $PF$ is slightly increased compared to the SL case (red bar), but is significantly reduced by $27\%$ compared to the pristine case (blue bar). We mention here that our simulations (not shown here), indicate that the degrading effect of energy relaxation can be prevented when OP emission is suppressed. This can be achieved by utilizing lower energies for transport compared to $\hbar \omega$ (lower Fermi level and consequently lower $V_{\text{SL}}$) or materials with large $\hbar \omega$, such that there is not enough energy range for emission to happen.

In a realistic scenario, however, the scattering is dominated by both elastic and inelastic processes. In the third column of Fig. 3 we show $G$, $S$, and $PF$, respectively, in the case in which AOP scattering is taken into account. Note that here the strengths of both scattering mechanisms are reduced to half of their initial values in order to have similar conductance numbers as in the AP case (for the pristine structure, blue bars) for a more reasonable comparison. Since the scattering rates and relaxation rates change, we cannot map quantitatively the results of this column to the previous two, but we treat it as a separate case and we only draw qualitative conclusions. In this scenario, the conductance in the case of SL barriers only (red bar) naturally increases in comparison to that of the pristine channel (blue bar) as the electrons propagate at higher velocity states (see Fig. 4(a)). Even if OP emission takes place, in this case the reduction in $G$ is not significant enough to lower it below that of the pristine channel. The Seebeck coefficient on the other hand, is reduced largely in the case of SL barriers in comparison to the pristine channel, which is similar to the case in which only OP scattering is considered (middle column). The origin of this similarity is that when both types of scattering processes are considered in this context, the energy relaxation is still determined by the OP scattering (note that as before we raise the $E_{\text{F}}$ in the SL channel). We also note in the inset of Fig. 3(c) that the average energy of the current flow measured from the Fermi level (i.e. Eq.~(\ref{eq14})), in the case of the pristine channel, $\langle E_{\text{pristine}} \rangle - E_F^{\text{pristine}} = 0.053$ eV (left), while in the case of a channel with SL barriers, $\langle E_{\text{SL}} \rangle - E_{\text{F}}^{\text{SL}} = 0.042$ eV (right), which is a factor of $\approx 0.79 \times$ smaller than that of the pristine case. Thus, similarly, the Seebeck coefficient shown in the bar chart drops from $S_{\text{pristine}} = 1.77 \times 10^{-4}$ V/K to $S_{\text{SL}} = 1.39 \times 10^{-4}$ V/K, i.e., by a factor of $\approx 0.78 \times$. Due to the large reduction in the Seebeck coefficient, the $PF$ is also degraded.

\textbf{Potential wells reduce $PF$ only slightly:} We now investigate the structures where the SL layers and the NIs introduce \textit{potential wells} for transport electrons (green-colored bars in Fig. 3). In the AP case, either of the two features reduce the conductance slightly, increase the Seebeck coefficient again very slightly, and thus the $PF$ suffers only slightly by $1.1\%$ and $2.6\%$, respectively, compared to the pristine material. As expected, potential wells cause some obstruction to transport due to reflections at the interfaces of the SL and NI boundaries, but this is not enough to cause significant reduction of the $PF$. The degradation of the $PF$ in the case in which inelastic scattering alone is taken into account is $11\%$ and $19\%$ for the SL wells structure and for the SL wells plus NIs structure, respectively (second column, third row of Fig.~3, compare blue vs green bars). This reduction is significantly less than that in the case of SL barriers and SL barriers plus NIs. This is due to the fact that, since the wells formed are thin in our case ($5$ nm), the carrier energy cannot relax easily in there, which therefore leads to a weak degrading influence.
\begin{figure}[t]
\vspace{-0in}
\hspace*{-0.3cm}
\includegraphics[width=8.6cm,height=10.4cm]{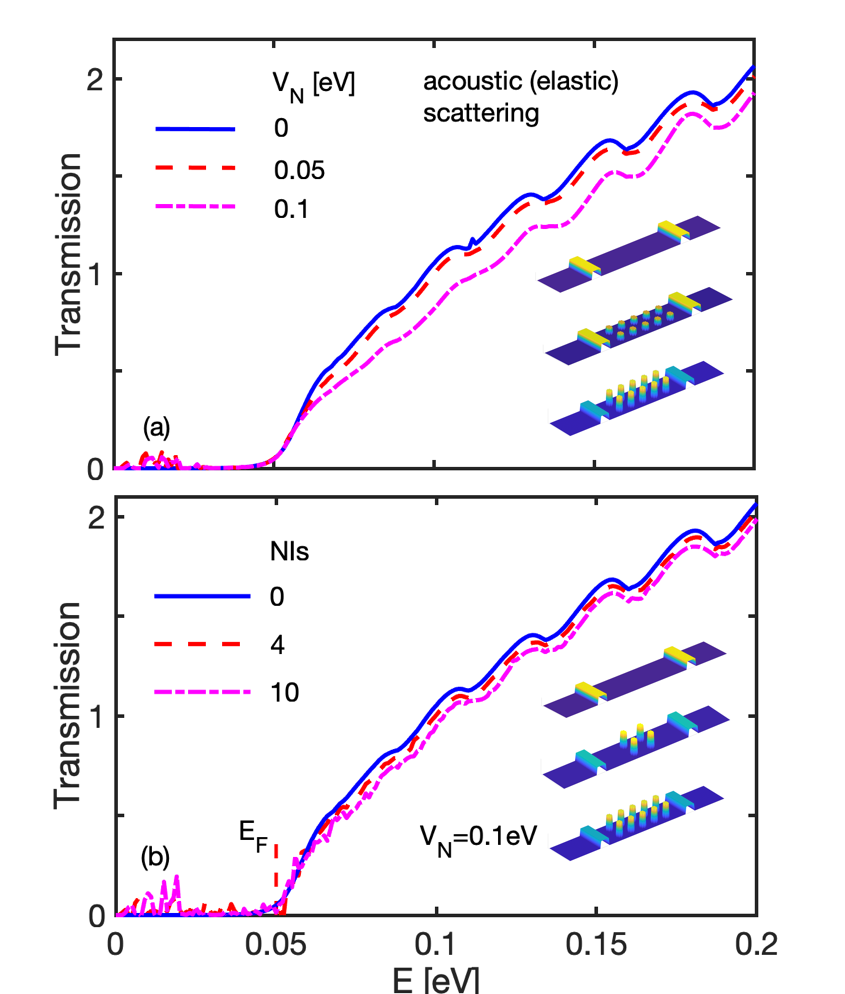}\\
\vspace*{0.2cm} \caption{\label{fig5} (Colour online)
Transmission $Tr$ versus electron energy $E$ for the channel with SL barriers and nanoinclusions (NIs) in the elastic scattering regime (AP scattering only) for: (a) increasing NI barrier height $V_{\text{N}}$, and (b) increasing number of NIs. In (b) the height of the NIs is set to $V_{\text{N}} = 0.1$ eV. The insets show schematics of the channels considered.
}
\end{figure}
\begin{figure*}
\vspace{-0in}
\hspace*{-0.6cm}
\includegraphics[width=19.2cm,height=12.5cm]{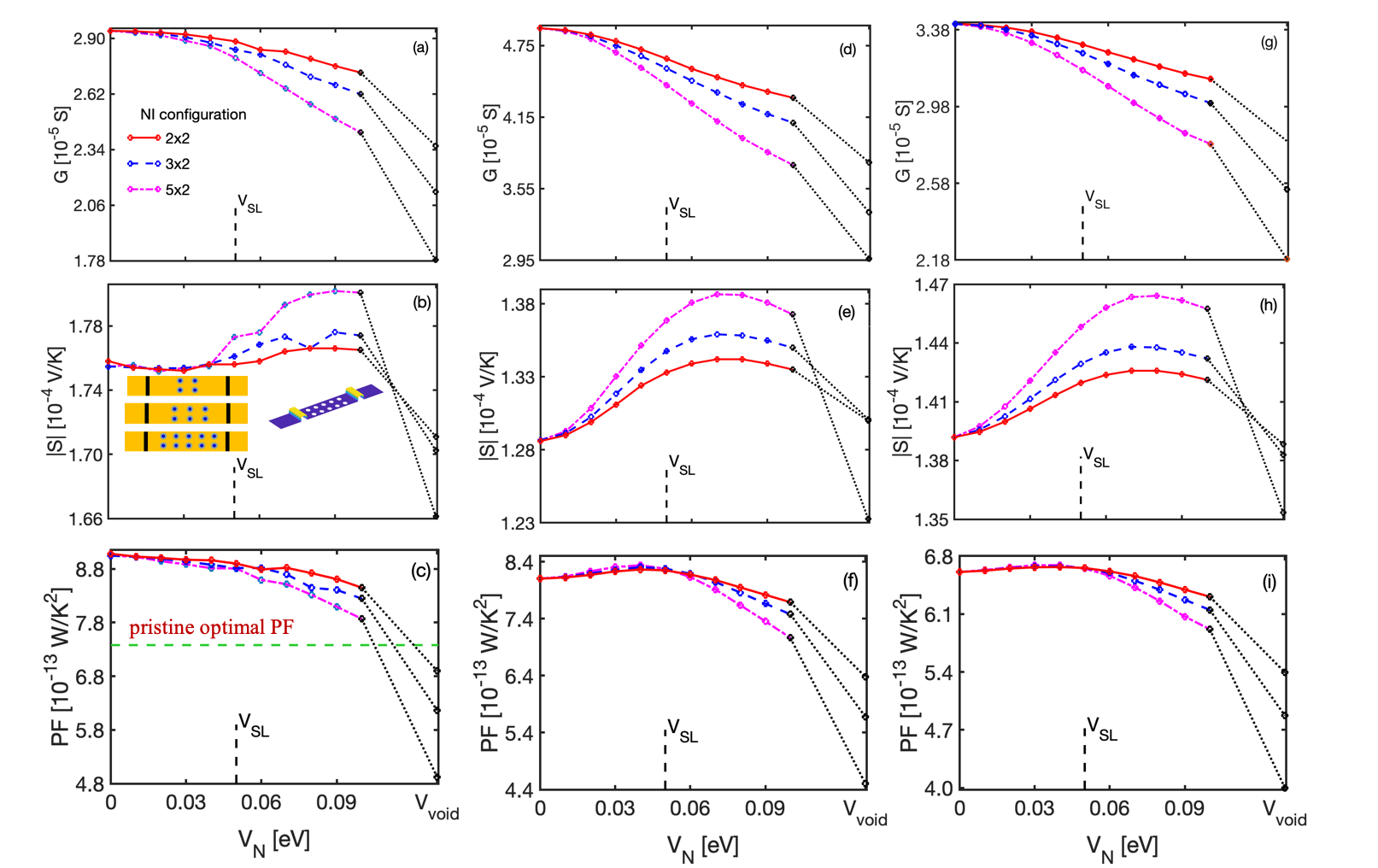}\\
\vspace*{0.2cm} \caption{\label{fig6} (Colour online)
Row-wise, summary of conductance $G$, Seebeck coefficient $S$, and power factor $PF$, for SL structures with barrier nanoinclusions (NIs) as shown in Fig. 1(c). Results for increasing NI barrier heights $V_{\text{N}}$ are presented. Column-wise, results for AP scattering only, OP scattering only, and AOP scattering are shown. In (c), the $PF$ of the optimized pristine channel (Fig. 1(a)) is shown by the green dashed line. The red, blue and magenta lines show results in which the number density of NIs increases from $2 \times 2$ to $3 \times 2$ and $5 \times 2$ NIs, respectively-as shown in the inset of (b). The dotted black lines extend the results to the case where the NIs are replaced with voids, as shown in the inset of (b) as well. The SL barrier height, $V_{\text{SL}}$, is denoted as well.
}
\end{figure*}

To summarize, therefore, in materials in which transport is dominated by elastic scattering, or if the inelastic scattering energy relaxation length is much larger that the characteristic geometrical features of the channel, it is beneficial to utilize nanostructures that form potential barriers, while setting high Fermi levels at the level of the SL barriers. In that case, benefits to the $PF$ by $>20\%$ can be achieved. In the case where the dominant scattering mechanisms are inelastic, then nanostructuring using potential wells is more beneficial. Although in this case improvements cannot be achieved, at least the reduction to the $PF$ is minimal. Later on we analyze these two cases in more detail.

\textbf{Robustness to $V_{\text{N}}$ and NI number density:} In the first case, where NIs form barriers, and under elastic scattering conditions, where benefits are observed, it is important to note that these benefits seem to be robust to the barrier heights of the NIs and their number density. Figure 5, for example, shows the transmission versus energy in the case where only AP scattering is considered for increasing values of $V_{\text{N}}$ (Fig. 5(a)) and for increasing number densities (Fig. 5(b)), as also illustrated in the insets. In Fig. 5a, by changing the barrier height from zero to $V_{\text{N}} = 0.05$ eV and then to $0.1$ eV, the transmission changes only slightly. The same is observed in Fig. 5(b) when the number of NIs changes from zero to $4$ and then to $10$. Minimal changes to the transmission are observed, indicating that the performance will be robust to such variations. It is known from a previous simulation work that the NI number density does not have a significant influence on the $PF$ when the Fermi level is raised in the bands and $V_{\text{NI}} \leq E_{\text{F}}$, because the moderate decrease they cause in $G$ is compensated by an increase in $S$ \cite{Foster17}. Experimental observations, where NIs embedded within matrix TE materials also point to this direction \cite{Fan11,Peng14,Zou14}. Here, in the hierarchical architecture, even better, the NI density does not affect the transmission, which means it affects neither $G$, nor $S$, nor the $PF$. The density, however, can drastically affect the thermal conductivity by increasing phonon scattering as shown in several works \cite{Biswas16, Biswas17,Shenghong18,Honarvar18,Yang18PRB}, which will benefit the overall $ZT$ figure of merit. We can attribute this difference in behavior to simple scattering theory, i.e. from Matthiessen’s rule the scattering mechanism with the smaller mean-free-path will have the largest effect. For example in Si the mean-free-path for electrons is of the order of few nanometers, but for phonons the dominant mean-free-path is $\sim 135$ nm - $300$ nm \cite{Sellan10,Dettori15}, which largely increases the influence of closely packed NIs on phonons, rather than electrons. 

\textbf{Comprehensive analysis:} In Fig. 6 we present a comprehensive analysis for the TE coefficients in the SL plus NIs barrier case,  as functions of the NI barrier height $V_{\text{N}}$ and for increasing NI number density. In a similar manner to Fig. 3, row-wise we show the TE coefficients $G$, $S$, and $PF$, while column-wise we show results for simulations that consider only AP scattering, only OP scattering, and AOP scattering, respectively. We consider the initial structure as the one which contains the SL barriers of height $V_{\text{B}} = 0.05$ eV, and we plot data versus the heights of the NI barriers $V_{\text{N}}$. In each sub-figure we show results for three structures, containing $4$, $6$ and $10$ NIs in the regions between the SL barriers (as shown in the inset of Fig. 6(b)). In all cases we observe that, as the NI barrier height increases, the conductance is reduced, however not strongly. The Seebeck coefficient demonstrates only a small increase, as the NIs tend to push $\langle E ( x ) \rangle$ slightly upward as shown in Fig. 2(b). Therefore, in the case of AP scattering in SL channels, the $PF$ exhibits a slight degradation of the order of $10\%$ when NIs are introduced (Fig. 6(c)). However, even at the high NI density and high $V_{\text{N}}$, the $PF$ is higher than that of the pristine material we began with (horizontal dashed-green line in Fig. 6(c)). A $\sim 10\%$ reduction in the $PF$ is observed in the case in which only OP scattering is considered (middle column, Fig. 6(f)), but in this case the $PF$ is already $\sim 30\%$ below the pristine material $PF$ value (not shown). In the third column of Fig. 6, where we consider the influence of AOP scattering, we observe again a $\sim4\%$ decrease in the $PF$ in comparison to the pristine case, an intermediate percentage value between the two extreme cases (although due to the scattering rates chosen, we cannot quantitatively compare this case to the previous two directly). In addition, as the number density of NIs increases, the conductance and $PF$ drop slightly, however their effect is not significant, even at high NI densities.

\textbf{The effect of voids:} The far right points connected by the black-dotted lines in the sub-figures of Fig. 6, indicate the corresponding results in the case where the NIs are replaced with voids. For simulation purposes, we increase $V_{\text{N}}$ in those geometries to very large numbers, effectively leading to vanishing wave function in those regions, which resembles a void structure. We notice that voids cause significant degradation in the conductance and in the $PF$; namely, there is $30\%$ - $50\%$ reduction from the SL reference depending on the NI number density (it turns out that in this case the density has a stronger effect). The Seebeck coefficient also seems to be reduced in the presence of voids, and interestingly it can be reduced to values below the Seebeck value of the SL channel without NIs or voids that we began with (left-most data points in Figs. 6(b), (e), and (h). The reasons behind this counter-intuitive simultaneous reduction in both conductance and Seebeck coefficient will be discussed later on. It is important to note, however, that voids degrade the thermal conductivity drastically, compared to NIs \cite{Dunham16,Taborda16}. Thus, despite the $\sim 50\%$ reduction in the $PF$, a large increase in the $ZT$ figure of merit is expected to be achieved in the void structures.

Note also that although in the presence of voids the conductance is degraded, the degradation is smaller when the Fermi level is placed at degenerate conditions, i.e., when $E_{\text{F}}$ is placed at the level of the SL barriers, in which case the carriers have higher velocities and are affected somewhat less. In Fig. 7(a) we show the transmission of the pristine channel and of the SL channel with/without voids, plotted versus the carrier energy. The blue/red lines correspond to the absence/presence of voids in a pristine channel (solid lines) and the channel with SL barriers (dashed lines), where the conduction opens after $V_{\text{SL}}$. The transmission in the SL case with or without voids (dashed lines), has a large slope after the energy crosses $V_{\text{SL}}$ (around the SL Fermi level), larger than the slope of the pristine channel (at $E=0$ eV, around the pristine channel Fermi level). In the pristine channel, upon the introduction of voids (red solid line), the transmission slope decreases significantly, starting and remaining close to zero for several meV above $E_{\text{F}}=0$ eV. Notice on the other hand, that in the SL case upon the introduction of voids (red-dashed line), the transmission is not degraded around $E = V_{\text{SL}}$ where the Fermi level is placed. As also seen earlier for the transmission of SL barriers in Fig.~5, this means that the conductance of the pore structures suffers less if we operate at degenerate conditions.
\begin{figure}[t]
\vspace{-0.35in}
\hspace*{-1.00cm}
\includegraphics[width=11cm,height=11.5cm]{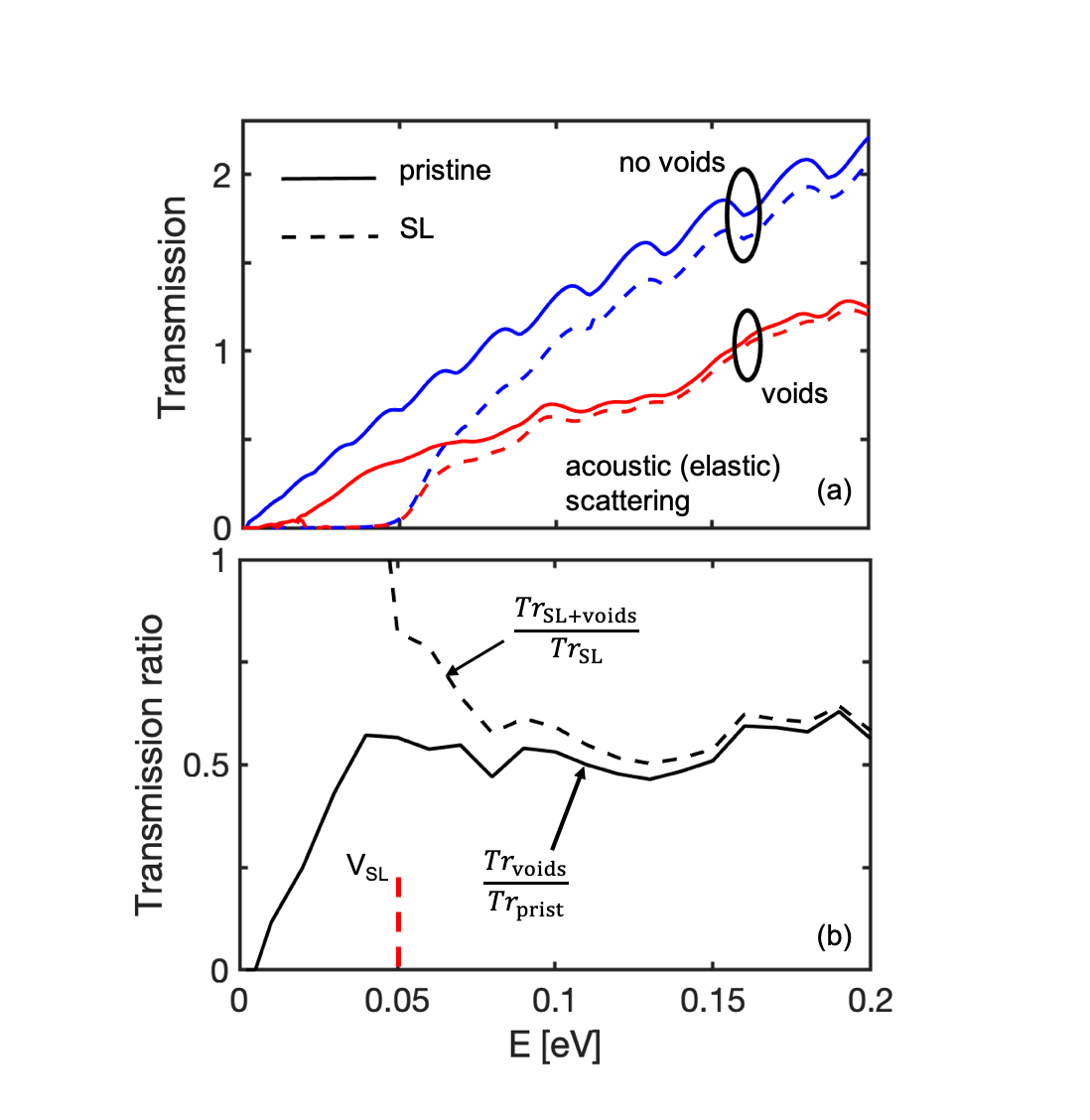}\\
\vspace*{-0.5cm} \caption{\label{fig7} (Colour online)
Comparison of the effect of voids on the transmission of the pristine channel (with $E_{\text{F}}$ at $0$ eV) and the SL channel, in which case the latter operates at highly degenerate conditions ($E_{\text{F}} = 0.05$ eV). (a) Transmission versus carrier energy of a pristine channel in the presence (red solid line) and in the absence (blue solid line) of voids. The dashed lines show the respective transmissions for a channel with SL barriers. (b) Transmission ratios $Tr_{\text{voids}} / Tr_{\text{prist}}$ and $Tr_{\text{SL+voids}} / Tr_{\text{SL}}$ plotted versus carrier energy.
}
\end{figure}
\begin{figure}[t]
\vspace{-0in}
\hspace*{-0.2cm}
\includegraphics[width=8.5cm,height=15cm]{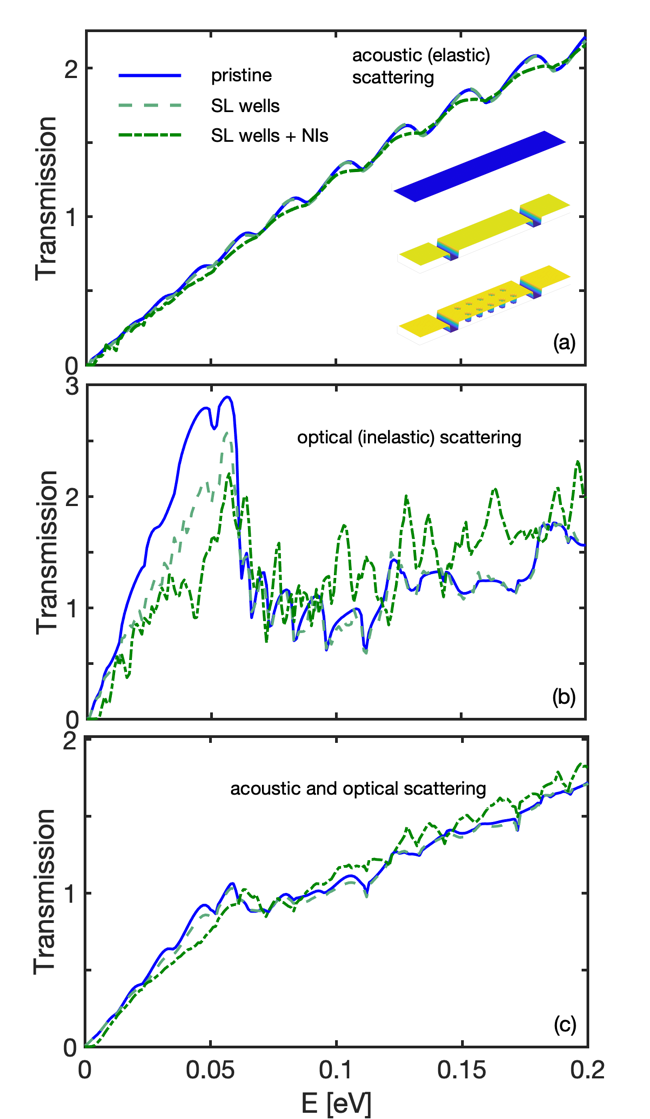}\\
\vspace*{-0cm} \caption{\label{fig8} (Colour online)
Transmission $Tr$ versus electron energy $E$ for the channel with SL wells and for the channel with SL wells plus NI wells between them for (a) elastic-AP scattering (b) inelastic-OP scattering, and (c) elastic/inelastic – AOP scattering regimes combined. The solid blue line shows results for the pristine channel, the green dashed line corresponds to the SL wells channel, and the green dashed-dotted line corresponds to the SL wells+NI channel, as indicated in the inset of (a).
}
\end{figure}

This is reflected more clearly in Fig. 7(b) where we show the ratio of the transmission of the pristine channel with voids to that of the pristine channel, $Tr_{\text{voids}} / Tr_{\text{prist}}$ (solid line) and the ratio of the transmission of the SL channels with voids to that of the SL channels, $Tr_{\text{SL+voids}} / Tr_{\text{SL}}$ (dashed line) plotted versus energy. In the SL channels the ratio of the transmissions starts from unity at $E = V_{\text{SL}}$, indicating the weak influence of the voids. In the pristine channel, on the other hand, the ratio begins at zero at $E=0$ eV, indicating that in this case the effect of voids is detrimental. This is quite important, indicating that the electronic conductance in highly disordered structures, which can slow down phonons significantly, can be less affected if they are operated at degenerate conditions, which will help the $PF$.

Making use of the observations in Fig.~7, we discuss briefly now the reason for which the values of the Seebeck coefficient of the SL channel with voids are smaller than those of the SL channel without voids (see Figs.~6(b), (e), and (h)). As can be observed in Fig.~7(a), for energies close to $V_{\text{SL}}$ (and also $E_\text{F}$) the SL+voids channel (red-dashed line) has similar transmission to that of the SL channel (blue-dashed line). However, at higher energies the transmission of the SL+voids channel grows with energy at a slower rate and it merges with the lower transmission of the pristine+voids channel. As a consequence the Seebeck coefficient, which is proportional to the slope of the transmission (in a similar manner that is proportional to the derivative of the density of states at the Fermi level), drops to lower values compared to those of the SL channel, i.e., the right-most points in Figs.~6(b), (e), and (h) are lower than the left-most points which correspond to $V_\text{N}=0$.

\textbf{Transport in the case of potential wells:} Finally, we examine the transport behavior of the structures in which the SL barriers and the NIs form potential wells. As indicated above in Fig. 3, the degradation in the conductance and in the $PF$ is very small, of the order of $\sim 1\%$. In Figs. 8(a)-(c) we show the transmission function versus energy in the cases where only AP scattering is present, only OP scattering is present, and where both elastic and inelastic scatterings are present, respectively. In each sub-figure we show three cases: i) pristine channel, ii) SL structure, and iii) SL structure plus NIs, all forming potential wells for electrons. These geometries are depicted in the insets of Fig. 8(a). Under elastic scattering conditions (Fig. 8(a)), in all three cases the transmission functions are almost identical, indicating that the nanostructuring does not obscure electronic transport. This is expected as the carriers flow at higher energies compared to the well energy levels. Some quantum reflections, however, are always present \cite{Thes15JAP,Foster17}, and thus some minor signatures are evident in the transmission features. In the case of OP scattering conditions in Fig. 8(b), a large peak is observed in the pristine case at low energies due to the lack of OP emission (blue line). At energies higher than the phonon energies $\hbar \omega = 0.06$ eV, a sharp drop is encountered, a result of the fact that the electrons have enough energy now to emit a phonon and lose energy all the way to the band edge. In the case of SL structures (green-dashed line) and the SL plus NIs geometries (green-dashed-dotted line), the initial peak is slightly suppressed, since now electrons have narrow regions (of the size of the SL and NI wells, $\sim 3$ nm - $5$ nm) to emit phonons and move to lower energies. However, this process is weak and the transmission is not significantly changed, which also shows why the conductance and $PF$ do not degrade noticeably compared to the pristine structure. In the case of Fig. 8(c), where AOP scattering is considered, the transmission function among the three geometries differs again slightly, the degree being intermediate between the completely elastic and completely inelastic behavior. This directly reflects the fact that the $PF$ does not change significantly from the pristine case when the wells are introduced. We note here, that our model considers nanostructuring as causing simple shifts in the band edges. In reality, however, the effective masses of the NIs will vary, strain fields will build around them, charging effects and interface resistances will appear, and the phonon-scattering details will also change. These will most probably add an additional reduction in the transmission which needs to be examined more carefully taking into account material specific parameters. However, our results demonstrate that to first order we should not expect large conductivity reductions from potential wells, as also is the case observed in experiments \cite{Zhao14}.

\section{Discussion on optimal nanostructuring for high power factors}

In our simulations we have chosen the simplest possible system that can account for energy filtering through barriers and energy relaxation through wells (i.e., only 2 barriers) without strong $PF$ degradation We did not intent to optimize the power factor of the hierarchical structure, it could be that higher power factors can be obtained for different geometries, but that would require significantly more work. Our purpose was to show that hierarchical architectures could provide benefits for the power factor, using geometrical features that we believe are close to optimal. Thus, here we elaborate on the reasons some of our choices are justified, which could also serve as useful guidelines to experimentalists in the design of high power factor advanced nanostructured thermoelectric materials in the presence of energy filtering:

i) \textit{$E_\text{F}$ and $V_\text{B}$ choice:} Degenerate conditions (with $E_F$ into the bands) are beneficial to employ high velocity electrons, and in that case the barrier heights $V_\text{B}$ and the Fermi level $E_\text{F}$ need to be positioned at similar levels. Energy filtering from potential barriers increases the Seebeck coefficient, however, not so easily the power factor unless these conditions are satisfied.

ii) \textit{Distance between filtering barriers:} This needs to be large enough because the more closely spaced the barriers are, the larger the resistance and the lower the conductivity, but on the other hand short enough, such that carriers do have space to relax their energy completely (this reduces the Seebeck coefficient and conductivity). Therefore, the distance is determined by the energy relaxation mean-free-path $\lambda_\text{E}$ (which is typically larger than the momentum relaxation mean-free-path). In our systems we have chosen an energy relaxation mean-free-path of $15.5$nm, and thus, a distance between the barriers of $50$nm, which is almost 3.5 times larger and allows semi-relaxation of the carrier energy. Indeed, for similar simulation scattering parameters, in Ref. \cite{Thes16JAP} we showed that the distance between barriers is optimized at around $50$nm.

iii) \textit{Spacing between the nanoinclusions (NIs) within}

\noindent\textit{the barriers:} From simple scattering mean-free-path considerations, it is common to assume that the NIs introduce a mean-free-path similar to the distance between them, and this needs to be combined with the momentum relaxation mean-free-path (not the energy relaxation-mean-free-path only), through Matthiessen’s rule. In our simulations, the distance between the NIs for the high density case is $d_{NI} = 6$nm, which is quite smaller compared to the mean-free-path of electron-phonon scattering. In principle, NIs degrade the power factor from its optimal value, and should be avoided if we only consider power factor improvements. However, they bring significant degradation in the thermal conductivity, for example in semiconductor materials for which the phonon mean-free path is $10$s-$100$s of nanometers. The important observation in this work, however, is that the degrading influence of NIs on the power factor of \textit{hierarchical architectures} is suppressed (at a larger degree compared to material cases that do not include the SL barriers). This is because the SL barriers (in combination with elevated Fermi levels) utilize charge carriers of higher energies, which are less susceptible to scattering from NIs. The scattering rate of high energy/large wavevector carriers by potential barriers is weaker, especially when the NI barrier height $V_{\text{NI}}$ is lower compared to the carrier energies (or negative in the case of wells). Thus, the recommendation for practical design of such hierarchical geometries is that the degrading effects on the power factor will be suppressed even if the NIs are placed at distances smaller compared to the mean-free-path of charge carriers.

With regards to improving the performance of thermoelectric materials, we need to mention here that although the electronic conductivity can be designed to be immune to the presence of NIs to a large degree, this is not the case for the thermal conductivity. A large number of literature reports indicate that NIs indeed cause significant degradation in the thermal conductivity \cite{Dettori15,Hahn_14, Neo_14JEM}. The combination of these two effects could decouple the electrical with the thermal conductivities and improve the $ZT$ figure of merit. There are two reasons why NIs affect phonons more than electrons: i) The distance between NIs can be thought of as the mean-free-path for scattering on the NIs. From simple Matthiessen’s rule scattering rate combination, the carrier with the longer mean-free-path will experience the larger relative reduction in its conductivity from a given NI geometry. Therefore, the thermal conductivity, carried by phonons with dominant mean-free-paths in the $10$s-$100$s of nanometers (in common semiconductors like Si), will experience a stronger reduction compared to the electronic conductivity, where electrons have mean-free-paths of a few to $10$s of nanometers. ii) Scattering of electrons on NIs is caused by the potential barriers that the NIs form. The electrons that contribute to conductivity are located energetically in a narrow window around the Fermi level, which can be shifted at high energies, where carriers are less obscured (especially if the barrier height $V_{\text{NI}}$ is small or negative). Phonon scattering on the other hand, does not offer this degree of freedom in the design of thermal conductivity. At room temperature for example, most phonons in the spectrum contribute to transport, and since phonons are lattice vibrations, they are affected by lattice interruptions. Therefore, although different nanostructuring can affect phonons with different mean-free-paths differently, all phonons are affected by NIs and the SL barriers.

\section{Summary and Conclusions}

In this work we investigated the influence of hierarchical nanostructuring on the thermoelectric coefficients of nanomaterials, which are the primer candidates for achieving ultra-low thermal conductivities and high thermoelectric $ZT$ figures of merit. Using the fully quantum mechanical NEGF transport formalism, we studied systematically two-dimensional materials with embedded SL-type barriers/wells combined with quantum dot-like NIs and voids. We found novel effects and presented design strategies for such materials, and stated the conditions under which the $PF$ is not only immune to nanostructuring, but it can also be improved. In summary, we showed that: 1) Nanostructuring using superlattice-like potential barriers and nanoinclusions can have up to $20\%$ $PF$ improvements even at very high nanoinclusion densities, as long as the Fermi level is placed well into the bands and charge carrier relaxation is avoided; 2) Nanostructuring using potential wells causes only minor reduction in the $PF$, even at very high nanostructuring densities. Thus, designing the nanostructured geometry of such materials should take into account the energy resolved mean-free-path of carriers, as well as their energy relaxation length caused by inelastic processes, in this case the optical phonon energies and the electron-optical phonon interaction strength. Such insight is currently not being explored in hierarchical nanostructured materials, where current strategies only focus on thermal conductivity reduction. It can, however, offer significant benefits to the thermoelectric  figure of merit by simultaneously improving, or at least not degrading the power factor as well.

{\bf Acknowledgments} This work has received funding from the European Research Council (ERC) under the European Union's Horizon 2020 Research and Innovation Programme (Grant Agreement No. 678763). We thank Dr Mischa Thesberg for helping with the construction of the NEGF simulator. We also thank Samuel Foster and Dhritiman Chakraborty for helpful discussions.

\appendix

\section{The acoustic deformation potential scattering rates}

For elastic acoustic deformation potential (ADP) scattering, where $\hbar \omega \rightarrow 0$, we use the commonly employed equipartition approximation. This results in the acoustic scattering rates to become proportional to the density of final states at the energy of the electronic state under consideration with the proportionality constant determined by the acoustic phonon deformation potential $D_{\text{A}}$, the temperature, and other material parameters. This process is described in detail in Ref. \cite{Lundstrom}, whereas the connection of the constant $D_{\text{AP}}$ used within NEGF to the actual deformation potential $D_{\text{A}}$ is presented in Ref. \cite{Koswatta07}. 

The ADP scattering rate is determined by \cite{Lundstrom}:
\begin{equation}
\frac{1}{\tau} = \frac{m^\ast D_{\text{A}}^2}{4 \pi \hbar \rho v_{\text{s}} p} \int_{\beta_{min}}^{\beta_{max}} 
\left ( n_{\text{B}} + \frac{1}{2} \mp \frac{1}{2} \right ) \beta^2 d \beta ,
\label{eqA1}%
\end{equation}
where $m^\ast$ is the effective mass of the material, $D_{\text{A}}$ is the deformation potential, $\rho$ is the mass density, $v_{\text{s}}$ is the sound velocity, and $p$ is the carrier momentum. $n_{\text{B}}$ is the number of phonons, determined by the Bose-Einstein distribution, and the integration is performed over all phonon wavevectors that participate in ADP scattering. Since the number of acoustic phonons at room temperature is large so that $n_{\text{B}} \simeq n_{\text{B}}+1$, and because $k_{\text{B}} T \gg \hbar \omega$, we can use the equipartition approximation $n_{\text{B}} \approx k_{\text{B}} T / \hbar \omega$. Taking also into account that the phonon dispersion is linear for acoustic phonons $(\omega = \upsilon_{\text{s}}\beta)$ the above equation can be simplified as
\begin{equation}
\frac{1}{\tau} = \frac{m^\ast D_{\text{A}}^2 k_{\text{B}} T}{2 \pi \hbar^2 \rho v_{\text{s}}^2 p} \int_{\beta_{min}}^{\beta_{max}} 
\beta d \beta ,
\label{eqA2}%
\end{equation}
where a factor of $2$ has been inserted due to both emission and absorption processes. In order to ensure momentum and energy conservation, $\beta_{min} = 0$ and $\hbar \beta_{max} = 2 p$, and therefore we get
\begin{equation}
\frac{1}{\tau} = \frac{m^\ast D_{\text{A}}^2 k_{\text{B}} T}{\pi \hbar^2 \rho v_{\text{s}}^2} \frac{k}{\hbar}  ,
\label{eqA3}%
\end{equation}
where $k = p / \hbar$ is the electron wave vector. Using $k = \left ( 2 m^\ast E \right )^{1/2} / \hbar$, and the density of states, $g(E)$, in 3D:
\begin{equation}
g(E) = \frac{1}{2 \pi^2} \left ( \frac{2 m^\ast}{\hbar^2} \right )^{3/2} E^{1/2} ,
\label{eqA4}%
\end{equation}
we can finally write the rate equation as:
\begin{equation}
\frac{1}{\tau} = \frac{\pi D_{\text{A}}^2 k_{\text{B}} T}{\hbar \rho v_{\text{s}}^2} g(E)  .
\label{eqA5}%
\end{equation}
Thus, the ADP scattering rate can be approximated with a constant times the density-of-states, which is standard practice.

\section{Seebeck coefficient as average energy of the current flow}

In order to extract the Seebeck coefficient $S$ we proceed as follows. In the Boltzmann transport formalism, the Seebeck coefficient is given by:
\begin{equation}
S = \frac{k_{\text{B}}}{q \sigma} \int_{-\infty}^{\infty} dE \left( - \frac{\partial f}{\partial E} \right) \Xi (E) \left( \frac{E - E_{F}}{k_B T} \right) ,
\label{eqB1}%
\end{equation} 
where $\Xi(E)$ is the transport distribution function. In terms of the energy and position resolved current $I_{ch}(E,x)$ the Seebeck coefficient can be expressed as
\begin{equation}
S^{\prime}(x) = \frac{1}{q I_{ch}} \int_{-\infty}^{\infty} dE I_{ch}(E,x)  \left( \frac{E - E_F}{T} \right) ,
\label{eqB2}%
\end{equation} 
which can be rewritten as
\begin{eqnarray}
\nonumber  S^{\prime}(x) = \frac{1}{q T} \left( \frac{1}{I_{ch}} \int_{-\infty}^{\infty} I_{ch}(E,x) E dE - E_F \right)
\\* &&\hspace*{-2.49in} = \frac{1}{qT} \langle E(x) - E_F \rangle  ,
\label{eqB3}%
\end{eqnarray}
where the integral in the last equation is the definition of the average energy of the current. The same derivation can be expressed in terms of the transmission by using the definition: $Tr_n(E) = (W/L) 2 \pi\hbar\Xi_n(E)$. The total Seebeck coefficient of the channel is then given as
\begin{equation}
S = \frac{1}{L} \int_{0}^{L} S^{\prime} ( x ) dx = \frac{1}{q T L} \int_{0}^{L} \langle E(x) - E_F \rangle dx ,
\label{eqB4}%
\end{equation}
where $q$ is the carrier charge ($q=-\vert e \vert$ for electrons and $q=\vert e \vert$ for holes) and $\langle E ( x ) \rangle$ is the energy of the current flow along the transport direction as given in Eq.~(\ref{eq14}).



\end{document}